\documentclass[12pt]{article} %[twocolumn]
\usepackage[top=30mm, bottom=40mm, left=25mm, right=25mm]{geometry}
\usepackage{cite} %for hyphened citations

\usepackage{graphicx}
\usepackage{amsmath, amssymb} 				% Needed for align environment
\usepackage{empheq}
\usepackage{fancybox}
\usepackage{mathrsfs} %for \mathscr
\usepackage{textcomp} % for \textinterrobang
\usepackage{mathtools}

\usepackage{epstopdf}

\usepackage{latexsym}
\usepackage{marvosym}
\usepackage{bm}	%This is for bold greek letter and such; command is \bm

\usepackage[mathcal]{euscript}
\usepackage{indentfirst}

\usepackage[rgb]{xcolor}

\usepackage{hyperref}
\hypersetup{
	unicode,	% use with \texorpdfstring
	colorlinks,
	citecolor=[rgb]{0,0,1},
	linkcolor=[rgb]{.5, 0, .5},
	urlcolor=[rgb]{.6,0,0},  % pdftex
	bookmarksopen=true,
	bookmarksopenlevel=\maxdimen,
	bookmarksnumbered
	}

\numberwithin{equation}{section}

\usepackage{amsthm} %Enable theorem environments
\theoremstyle{definition}

\theoremstyle{plain}

\font\biggest=cmssbx10 scaled 3200

\def\K3{\mathrm K3}

\def\gl#1#2{$\mathrm{GL}(#1; {\bf #2})$}
\def\sl#1#2{$\mathrm{SL}(#1; {\bf #2})$}
\def\sp#1#2{$\mathrm{Sp}(#1; {\bf #2})$}
\def\spin#1#2{$\mathrm{Spin}(#1, #2)$}
\def\su#1#2{SU({#1,#2})}
\def\usp#1#2{USp({#1,#2})}
\def\U#1{U({#1})}

\def\double #1{#1{\hbox{\kern-2pt $#1$}}}
\def\un#1{\underline #1}
\def\gl#1#2{\ifmmode \mathrm{GL}(#1; {\bf #2}) \else $\mathrm{GL}(#1; {\bf #2})$\fi}
\def\sl#1#2{\ifmmode \mathrm{SL}(#1; {\bf #2}) \else $\mathrm{SL}(#1; {\bf #2})$\fi}
\def\so#1{\ifmmode \mathrm{SO}({#1}) \else $\mathrm{SO}(#1)$\fi}

\def\sp#1#2{\ifmmode \mathrm{Sp}(#1; {\bf #2}) \else $\mathrm{Sp}(#1; {\bf #2})$\fi}
\def\usp#1#2{\ifmmode \mathrm{USp}(#1,#2) \else $\mathrm{USp}(#1,#2)$\fi}
\def\spin#1#2{\ifmmode \mathrm{Spin}(#1,#2) \else $\mathrm{Spin}(#1,#2)$\fi} 
\def\su#1{\ifmmode \mathrm{SU}({#1}) \else $\mathrm{SU}(#1)$\fi}

\let\sec=\S

\def\on#1#2{{\buildrel{\mkern2.5mu#1\mkern-2.5mu}\over{#2}}}
\def\dt#1{\on{\hbox{\bf .}}{#1}}                % (big) dot over: seeÀ   below  
\mathcode`\*="702A                  % * now always complex conjugate
\def\f#1#2{{\textstyle{#1\over#2}}}	   % fraction
\def\half{{\textstyle{1\over{\raise.1ex\hbox{$\scriptstyle{2}$}}}}}
\def\slap#1#2{\setbox0=\hbox{$#1{#2}$}#2\kern-\wd0{\hbox to\wd0{\hfil$#1{/}$\hfil}}}
\def\sla#1{\mathpalette\slap{#1}}		%slash: see ÖÊÊÊbelow

\def\Gamma{\mathchar"0100}
\def\Delta{\mathchar"0101}
\def\Theta{\mathchar"0102}
\def\Lambda{\mathchar"0103}
\def\Xi{\mathchar"0104}
\def\Pi{\mathchar"0105}
\def\Sigma{\mathchar"0106}
\def\Upsilon{\mathchar"0107}
\def\Phi{\mathchar"0108}
\def\Psi{\mathchar"0109}
\def\Omega{\mathchar"010A}

\catcode128=13 \def €{\"A}                 % Option-u A
\catcode129=13 \def {\AA}                 % Option-A
\catcode130=13 \def '{\c}           	   % Option-C (cedilla)
\catcode131=13 \def ƒ{\'E}                   % Option-e E
\catcode132=13 \def "{\~N}                   % Option-n N
\catcode133=13 \def …{\"O}                 % Option-u O
\catcode134=13 \def †{\"U}                  % Option-u U
\catcode135=13 \def ‡{\'a}                  % Option-e a
\catcode136=13 \def ˆ{\`a}                   % Option-`  a
\catcode137=13 \def ‰{\^a}                 % Option-i a
\catcode138=13 \def Š{\"a}                 % Option-u a
\catcode139=13 \def ‹{\~a}                   % Option-n a
\catcode140=13 \def Œ{\alpha}            % Option-a
\catcode141=13 \def {\chi}                % Option-c
\catcode142=13 \def Ž{\'e}                   % Option-e e
\catcode143=13 \def {\`e}                    % Option-`  e
\catcode144=13 \def {\^e}                  % Option-i e
\catcode145=13 \def '{\"e}                % Option-u e
\catcode146=13 \def '{\'\i}                 % Option-e i
\catcode147=13 \def "{\`\i}                  % Option-`  i
\catcode148=13 \def "{\^\i}                % Option-i i
\catcode149=13 \def •{\"\i}                % Option-u i
\catcode150=13 \def –{\~n}                  % Option-n n
\catcode151=13 \def —{\'o}                 % Option-e o
\catcode152=13 \def ˜{\`o}                  % Option-`  o
\catcode153=13 \def ™{\^o}                % Option-i o
\catcode154=13 \def š{\"o}                 % Option-u o
\catcode155=13 \def ›{\~o}                  % Option-n o
\catcode156=13 \def œ{\'u}                  % Option-e u
\catcode157=13 \def {\`u}                  % Option-`  u
\catcode158=13 \def ž{\^u}                % Option-i u
\catcode159=13 \def Ÿ{\"u}                % Option-u u
\catcode160=13 \def  {\tau}               % Option-t
\catcode162=13 \def ¢{\oplus}           % Option-4
\catcode163=13 \def £{\relax\ifmmode\expandafter\to\else\expandafter\itemize\fi} % Option-3
\catcode164=13 \def ¤{\subset}	  % Option-6
\catcode165=13 \def ¥{\infty}           % Option-8
\catcode166=13 \def ¦{\mp}                % Option-7
\catcode167=13 \def §{\sigma}           % Option-s
\catcode168=13 \def ¨{\rho}               % Option-r
\catcode169=13 \def ©{\gamma}         % Option-g
\catcode170=13 \def ª{\leftrightarrow} % Option-2 ; Option-E (acute) :
\catcode171=13 \def «{\relax\ifmmode\expandafter\acute\else\expandafter\'\fi}
\catcode172=13 \def ¬{\relax\ifmmode\expandafter\ddt\else\expandafter\"\fi}
\catcode173=13 \def ­{\equiv}            % Option-= ; ^ Option-U (umlaudt)
\catcode174=13 \def ®{{{}={}}}          % Option-"
\catcode175=13 \def ¯{\Omega}          % Option-O
\catcode176=13 \def °{\otimes}          % Option-5
\catcode177=13 \def ±{\ne}                 % Option-+
\catcode178=13 \def ²{\le}                   % Option-,
\catcode179=13 \def ³{\ge}                  % Option-.
\catcode180=13 \def ´{\upsilon}          % Option-y
\catcode181=13 \def µ{\mu}                % Option-m
\catcode182=13 \def ¶{\delta}             % Option-d
\catcode183=13 \def ·{\epsilon}          % Option-w
\catcode184=13 \def ¸{\Pi}                  % Option-P
\catcode185=13 \def ¹{\pi}                  % Option-p
\catcode186=13 \def º{\beta}               % Option-b
\catcode187=13 \def »{\partial}           % Option-9
\catcode188=13 \def ¼{\nobreak\ }       % Option-0
\catcode189=13 \def ½{\zeta}               % Option-z
\catcode190=13 \def ¾{\sim}                 % Option-'
\catcode191=13 \def ¿{\omega}           % Option-o
\catcode192=13 \def À{\dt}                     % Option-?
\catcode193=13 \def Á{\gets}                % Option-1
\catcode194=13 \def Â{\lambda}           % Option-l
\catcode195=13 \def Ã{\nu}                   % Option-v
\catcode196=13 \def Ä{\phi}                  % Option-f
\catcode197=13 \def Å{\xi}                     % Option-x
\catcode198=13 \def Æ{\psi}                  % Option-j
\catcode199=13 \def Ç{\int}                    % Option-\
\catcode200=13 \def È{\oint}                 % Option-|
\catcode201=13 \def É{\relax\ifmmode\expandafter\cdot\else\vol\fi}    % Option-;
\catcode202=13 \def Ê{\relax\ifmmode\expandafter\,\else\thinspace\fi}
\catcode203=13 \def Ë{\`A}                      % Option-`  A ; ^ Option-space
\catcode204=13 \def Ì{\~A}                      % Option-n A
\catcode205=13 \def Í{\~O}                      % Option-n O
\catcode206=13 \def Î{\Theta}              % Option-Q
\catcode207=13 \def Ï{\theta}               % Option-q; Option-- :
\catcode208=13 \def Ð{\relax\ifmmode\expandafter\bar\else\expandafter\=\fi}
\catcode209=13 \def Ñ{\overline}             % Option-_
\catcode210=13 \def Ò{\langle}               % Option-[
\catcode211=13 \def Ó{\relax\ifmmode\expandafter\{\else\ital\fi}      % Option-{
\catcode212=13 \def Ô{\rangle}               % Option-]
\catcode213=13 \def Õ{\}}                        % Option-}
\catcode214=13 \def Ö{\sla}                      % Option-/; Option-V :
\catcode215=13 \def ×{\relax\ifmmode\expandafter\check\else\expandafter\v\fi}
\catcode216=13 \def Ø{\"y}                     % Option-u y
\catcode217=13 \def Ù{\"Y}  		    % Option-u Y
\catcode218=13 \def Ú{\Leftarrow}       % Option-!
\catcode219=13 \def Û{\Leftrightarrow}       % Option-@ ; Option-# :
\catcode220=13 \def Ü{\relax\ifmmode\expandafter\Rightarrow\else\sect\fi}
\catcode221=13 \def Ý{\sum}                  % Option-$
\catcode222=13 \def Þ{\prod}                 % Option-%
\catcode223=13 \def ß{\widehat}              % Option-^
\catcode224=13 \def à{\pm}                     % Option-&
\catcode225=13 \def á{\nabla}                % Option-(
\catcode226=13 \def â{\quad}                 % Option-)
\catcode227=13 \def ã{\in}               	% Option-W
\catcode228=13 \def ä{\star}      	      % Option-R
\catcode229=13 \def å{\sqrt}                   % Option-M
\catcode230=13 \def æ{\^E}			% Option-i E
\catcode231=13 \def ç{\Upsilon}              % Option-Y
\catcode232=13 \def è{\"E}    	   	 % Option-u E
\catcode233=13 \def é{\`E}               	  % Option-`  E
\catcode234=13 \def ê{\Sigma}                % Option-S
\catcode235=13 \def ë{\Delta}                 % Option-D
\catcode236=13 \def ì{\Phi}                     % Option-F
\catcode237=13 \def í{\`I}        		   % Option-`  I
\catcode238=13 \def î{\iota}        	     % Option-H
\catcode239=13 \def ï{\Psi}                     % Option-J
\catcode240=13 \def ð{\times}                  % Option-K
\catcode241=13 \def ñ{\Lambda}             % Option-L
\catcode242=13 \def ò{\cdots}                % Option-:
\catcode243=13 \def ó{\^U}			% Option-i U
\catcode244=13 \def ô{\`U}    	              % Option-`  U
\catcode245=13 \def õ{\bo}                       % Option-B ; Option-I :
\catcode246=13 \def ö{\relax\ifmmode\expandafter\hat\else\expandafter\^\fi}
\catcode247=13 \def÷{\relax\ifmmode\expandafter\tilde\else\expandafter\~\fi}
\catcode248=13 \def ø{\ll}                         % Option-< ; ^ Option-N
\catcode249=13 \def ù{\gg}                       % Option-> 
\catcode250=13 \def ú{\eta}                      % Option-h 
\catcode251=13 \def û{\kappa}                  % Option-k 
\catcode252=13 \def ü{\half}     		 % Option-Z 
\catcode253=13 \def ý{\Gamma} 		% Option-G 
\catcode254=13 \def þ{\Xi}   			% Option-X ; Option-T : 
\catcode255=13 \def ÿ{\relax\ifmmode\expandafter{}^{\dagger}{}\else\dag\fi}

\def\A{{\cal A}}  \def\B{{\cal B}}    
         
  \def\K{{\cal K}}         
  \def\P{{\cal P}}    
\def\S{{\cal S}}  \def\T{{\cal T}}  \def\U{{\cal U}} \def\V{{\cal V}}  \def\W{{\cal W}}
\def\X{{\cal X}}    

\def\^{\wedge}

\def\dd{\hbox{\,\Large$\triangleright$}}

\def\dig#1{\setbox0=\hbox{$#1M$}
	\hskip.06\wd0 \vrule width.07\wd0 height.63\wd0 depth.01\wd0 
	\vrule width.37\wd0 height.63\wd0 depth-.56\wd0 \hskip-.4\wd0
	\vrule width.25\wd0 height.35\wd0 depth-.28\wd0 
	\vrule width.07\wd0 height.35\wd0 depth-.17\wd0 \hskip.14\wd0}
\def\digamma{{\mathpalette\dig{}}}
\def\di{\digamma}

\title{
{\color{green}{
\biggest
F-brane Dynamics}}
}
\author{William D. Linch \textsc{iii}$\,{}^\text{\Pisces}$ and Warren Siegel$\,{}^\text{\Scorpio}$}
\date{{\color{red} 
}}		% Activate to display a given date or no date

\begin{document}
\maketitle

%email addresses
\vspace*{-75mm}
\begin{flushright}
{MI-TH-1628\\
YITP-SB-16-38}
\end{flushright}
\vspace*{+50mm}

\begin{center}
%\vskip 0.2in
{\em
${}^{\mbox{\footnotesize\Pisces}}$
%George P. and Cynthia W. 
Mitchell Institute for
Fundamental Physics and Astronomy, \\
Texas A\&{}M University,
College Station, TX 77843\\
~\\
${}^{\mbox{\footnotesize\Scorpio}}$
C. N. Yang Institute for Theoretical Physics\\
State University of New York, Stony Brook, NY 11794-3840
}%\\
\end{center} 

\vspace{10pt}

\begin{abstract}
We generalize the current algebra of constraints of U-duality-covariant critical superstrings to include the generator responsible for the dynamics of the fundamental brane. 
This allows us to define $\kappa$ symmetry and to write a worldvolume action in Hamiltonian form that is manifestly supersymmetric in the target space. 
The Lagrangian form of this action is generally covariant, but the worldvolume metric has fewer components than expected. 
\end{abstract}

%email addresses
\vspace*{.5cm}
\begin{flushleft}
~\\
{${}^{\mbox{\footnotesize\Pisces}}$ \href{mailto:wdlinch3@gmail.com}{wdlinch3@gmail.com}}\\
%\makebox[0pt][t]
{$^{\text{\Scorpio}}$ \href{mailto:siegel@insti.physics.sunysb.edu}{siegel@insti.physics.sunysb.edu}}
%\makebox[0pt][t]
\end{flushleft}

\setcounter{page}1
\thispagestyle{empty}
\newpage

%\setcounter{page}0
%\thispagestyle{empty}
%{\linespread{1}
\tableofcontents
%}

\newpage
%%%%%%%%%%%%%%%%%%%%%%%%%%%%%%%%%%%%%%%%%%%%%%%%%%%%%%%%%%%%%%%%
%%%%%%%%%%%%%%%%%%%%%%%%%%%%%%%%%%%%%%%%%%%%%%%%%%%%%%%%%%%%%%%%
\section{Introduction}
\label{S:Introduction}

Critical superstring theories are related to each other by U-duality. 
(We think of T-duality as compactifying, dualizing, and decompactifying). 
M-theory \cite{Witten:1995ex} may be thought of as the orbit of such theories under this U-duality action. 
Roughly speaking, the fundamental objects of M-theory analogous to strings are related to M2-branes, M5-branes, {\it et cetera}, but U-duality mixes these up. 
Contrary to the situation in string theory, however, M2's and their dual M5's are simultaneously light. 
This suggests that the fundamental object to quantize in M-theory should contain all the simultaneously ``light'' branes in the same U-duality orbit. 
In particular, we should not try to quantize membranes alone.

An alternative is to quantize forms on fundamental branes to obtain manifestly U-duality-covariant current algebras \cite{Linch:2015fya, Linch:2015qva, Linch:2015fca}. 
(This is a natural extension to higher-dimensional worldvolumes of what is done for the string, as we review in \sec{}\ref{S:CurrentAlgebra}.)
These are the fundamental currents of a formalism we call F-theory.\footnote{Usually the term ``M-theory" is used in a way that is too vague, and ``F-theory" in a way that is too specific.  
We use explicit constructions for both, relating them to dynamical, fundamental branes.  
In particular, our definition of F-theory corresponds to the original one: In reference \cite{Vafa:1996xn}, Vafa specifically proposes a generalization/reformulation of (string theory and) M-theory to higher dimensions (cf.\ \sec{}3 of \cite{Vafa:1996xn} in particular).
Confusion about this point seems to have arisen because it is now common practice to use ``F-theory" to refer to what was originally called ``evidence for F-theory" in \cite{Vafa:1996xn}.
%The relation to (as explained in detail for the rank-1 case in \cite{Berman:2015rcc})
} %end footnote

Insight into the interpretation of these fundamental currents can be obtained by comparing to the ``effective'' brane current algebras resulting from the canonical analysis of the standard M2/M5 embeddings
%. These M2/M5-brane models and their actions were 
originally constructed in 
\cite{Bergshoeff:1987cm, Bergshoeff:1987qx, Schwarz:1993vs, Howe:1996mx, Howe:1996yn, Perry:1996mk, Howe:1997fb, Sezgin:1998tm}.\footnote{Pasti and collaborators give an equivalent description of these with an auxiliary gauge field \cite{Pasti:1997gx, Bandos:1997ui}. Unfortunately, such reformulations are not any more covariant, since there is no freedom to choose inequivalent gauges and thus {\em must} reduce to the original theory. (This is analogous to the fact that one can always make non-relativistic mechanics look ``manifestly relativistic'' by including compensating fields for the missing Poincar\'e generators even though the theory does not have this symmetry.)
} %end footnote
Hatsuda and Kamimura have carried out the canonical analysis of these actions for the classical M2 \cite{Hatsuda:2012vm} and M5 branes \cite{Hatsuda:2013dya}.
They obtain effective currents in the same representations as the fundamental currents of \cite{Linch:2015fya, Linch:2015qva, Linch:2015fca}. 
The former are complicated composite currents given in terms of the embedding coordinates, background fields, and so on, which can be shown \cite{HatsudaPrivate} to satisfy the same algebra as the fundamental currents of the $E_{5(5)} = Spin(5,5)$ theory of reference \cite{Linch:2015qva}.
This comparison suggests that our fundamental $E_{5(5)}$ currents describe simultaneously the momentum, M2 charge, and M5 charge combined into the 16-dimensional representation of $E_5$. 

\begin{figure}[t]
$$ \vcenter{\hbox{\includegraphics[width=3.5in]{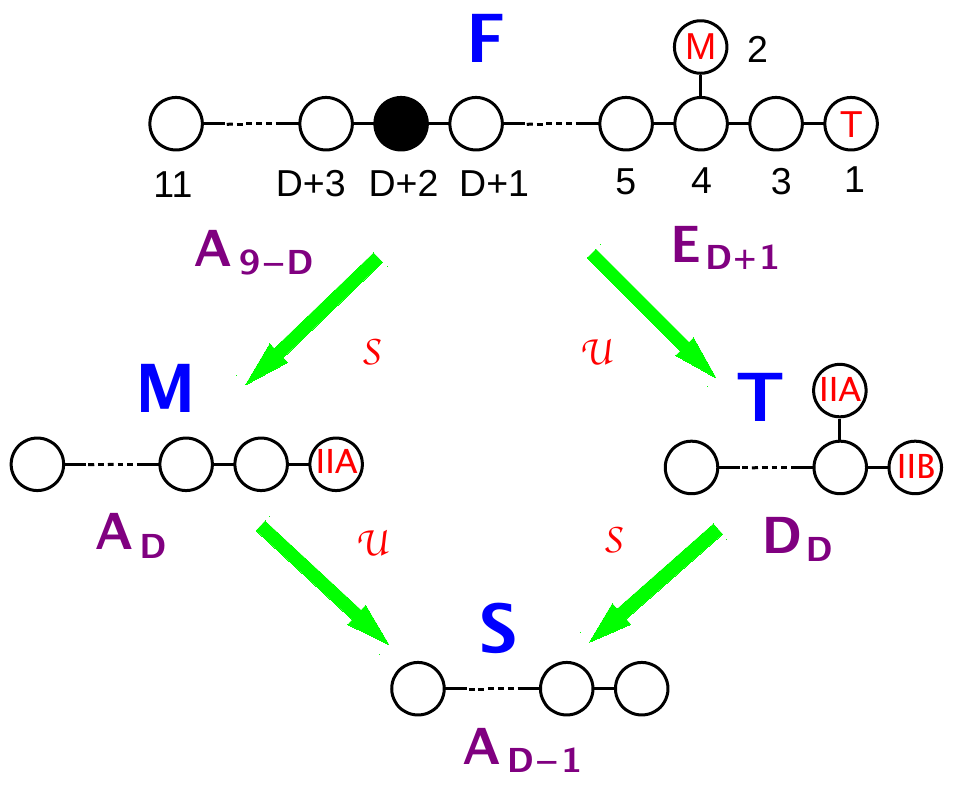}}} $$
\caption{\small The F-theory diamond. The black circle is eliminated to divide the symmetry group $E_{11}$ into the ``spacetime" symmetry $E_{{\mathrm D}+1}$ acting on the $X$'s and the ``internal" one $A_{9-{\mathrm D}}$ acting on the $Y$'s.  The circles are labeled in red to indicate which theories are produced when they are removed via reduction, by the constraints labeling the arrows.  The Dynkin diagrams and spacetime symmetries for the reduced theories are indicated. (All algebras are in their split forms.)}
\label{F:Fdiamond}
\end{figure}

We emphasize that we are not advocating the consistency of the effective/probe M-brane theory (see {\it e.g.\ }\cite{Duff:2015jka}). Instead, we propose to replace it with a fundamental, quantizable theory that generates an isomorphic current algebra. 
Such theories are manifestly $E_{n(n)}$-covariant descriptions of critical superstrings in which ${\mathrm D}=n-1$ of the 10 dimensions exhibit manifest super-Poincar\'e invariance. 
(Ultimately, $n$ should be taken up to 11 \cite{Duff:1985bv, Duff:1986ne, West:2001as, Tumanov:2015yjd, Tumanov:2016abm}, but only those with $n\leq7$ have been constructed to date \cite{Linch:2015fca}.)
These theories can be represented in terms of the $E_{11}$ Dynkin diagram in the top row of figure \ref{F:Fdiamond}.
In this representation, the black node is eliminated to divide the symmetry group $E_{11}$ into the ``spacetime" symmetry $E_{{\mathrm D}+1}$ acting on coordinates we will call $X$ and the ``internal" $A_{9-{\mathrm D}}= SL(10-{\mathrm D})$ symmetry acting on the remaining $10-{\mathrm D}$ coordinates $Y$.

Closure of the algebra of Virasoro-like currents $\S$ of these F-theories 
requires a Gau\ss{} law constraint $\U$. 
(This constraint arises since we obtain these theories by quantizing gauge $p$-forms.)
It also requires ``strong sectioning'': solving $\S=0$ on the tensor algebra of the Hilbert space. Doing this spontaneously breaks $E_{{\mathrm D}+1} \to GL({\mathrm D}+1)$, the symmetry group of $({\mathrm D}+1)$-dimensional M-theory. 
(Abelian factors GL(1) do not appear in Dynkin diagrams.)
Alternatively, solving $\U=0$ reduces $E_{{\mathrm D}+1} \to SO({\mathrm D},{\mathrm D})$ and the manifestly T-duality-covariant description of the ${\mathrm D}$-dimensional type-II string \cite{Siegel:1993xq, Siegel:1993th, Siegel:1993bj} is recovered. 
Solving both reduces $E_{{\mathrm D}+1} \to GL({\mathrm D})$ of S(tring)-theory.

The Virasoro-like constraints $\S$ of the $E_{n(n)}$ theory generate translations on the worldvolume \cite{Linch:2015fya} (see also eq.\ \ref{E:wvdiff} below).
Since they also transform in a fundamental representation of $E_{n(n)}$, the target space symmetries are mixed up with the worldvolume ones. 
To be more explicit, let us denote the target space Lorentz group by $G$ $(=E_{n(n)})$ and the worldvolume Lorentz group by $L$. 
As much of the analysis is canonical, the basic worldvolume symmetry is Hamiltonian. We refer to this formalism and its structure group interchangeably as ``$H$''. Thus, the canonical analysis reduces $L \to H$ and the statement becomes that the $G$-covariant constraints $\S$ transform in the defining representation of $H$. 
This is only possible if $G$ has a subgroup that is not only isomorphic to $H$ but is, in fact, identified with it.
%\footnote{This may be thought of as twisting: Let $H_L\subset L$ and $H_G\subset G$ denote the relevant isomorphic subgroups. We may think of $H_L$ as the Lorentz version of $H$ and $H_G$ as the ``R-symmetry'' version of it. Twisting by $H_G$ is usually taken to mean that the $H$ group of the twisted theory is the diagonal subgroup of $H_L \times H_G$.
%} % end footnote

For low ranks, we have shown that these theories arise from the canonical analysis of selfdual $p$-forms on certain branes \cite{Linch:2015fya, Linch:2015qva}. Such descriptions have Lagrangian symmetry, and we again refer to both this formulation of the theory and its underlying group as ``$L$''. Then the group structures can be summarized as $G \supset H \subset L$.
In this article, we extend the constraint algebra by including the Virasoro operators $\T$ associated with $\tau$-development ($\S^a$ generates only the translations in $\sigma_a$; cf.\ eq.\ \ref{E:wvdiff}) and use it to study the $L$(agrangian) formulation of our theories. 

We conclude this introduction with an outline of our presentation. 
We begin by extending the worldvolume constraint algebra in the $H$(amiltonian) formalism in section \ref{S:CurrentAlgebra}.
Currents and constraints for the bosonic theory are reviewed in \ref{S:Constraints}, and their algebra is worked out in section \ref{S:Algebra}. In section \ref{S:Brackets}, we review the derivation of the D- and C-brackets (the F-theory generalizations of the exceptional Lie derivative and bracket on the target). 

Closure of the constraint algebra requires the Gau\ss{} law and Laplace constraints $\U$ and $\V$. 
We give an interpretation of these constraints in section \ref{S:F-orms} as an F-theory generalization of de Rham forms, or ``F-orms'' for brevity. 
In particular, in section \ref{S:HFS}, the $H$-covariant form of the field strengths is given in terms of differentials related to the Gau\ss{} constraint. 
This and the Laplace constraint are studied in sections \ref{S:UOnly} and \ref{S:UV} where they are related to the existence of a differential on a complex of F-orms consisting of the gauge parameters, fundamental fields, field strengths, Bianchi identities, and so on. This complex is constructed explicitly in section \ref{S:LFS} in terms of $L$-covariant field strengths.

In section \ref{S:Critical}, we complete this bosonic F-theory to the critical superstrings by introducing the scalars and Green-Schwarz fermions. 
We define the $\kappa$ symmetry generators $\B^{\un \alpha}$ corresponding to the first-class part of the spinorial constraints $D_{\un \alpha}$. 
We then investigate the $L$(agrangian) description of the theory by performing a Legendre transformation on the $H$(amiltonian) action in section \ref{S:Action}. 
The result is manifestly supersymmetric in the target space but has a peculiar structure on the worldvolume:
Although it is covariant by construction, the worldvolume metric has only d (= dimension of worldvolume)
%${\mathrm d}=\mathrm{dim}(\Sigma)$ 
independent components instead of the expected $\tfrac12{\mathrm d}({\mathrm d}+1)$. 
Such Lagrangians are studied in section \ref{S:ActionsSD} by constructing them from selfdual F-orms. They have Wess-Zumino terms of ``heterotic'' type (no $\Theta^4$ term). 
Finally, we reduce these actions from F-theory to T-theory (\sec{}\ref{S:F2T}) by solving the Gau\ss{} law sectioning condition thereby showing how the type-II Wess-Zumino term is generated. 

Our conclusions are reviewed in section \ref{S:Conclusion}.
We include 
three appendices 
\ref{S:Notation}, 
\ref{S:SummarySymmetry}, and  
\ref{S:eta}
%respectively 
summarizing our notation, the symmetry structure of the theories, and the explicit form of the Clebsh-Gordan-Wigner tensor defining the fundamental currents for each dimension.

%%%%%%%%%%%%%%%%%%%%%%%%%%%%%%%%%%%%%%%%%%%%%%%%%%%%%%%%%%%%%%%%
%%%%%%%%%%%%%%%%%%%%%%%%%%%%%%%%%%%%%%%%%%%%%%%%%%%%%%%%%%%%%%%%
\section{Current Algebra}
\label{S:CurrentAlgebra}
Consider the theory of a chiral $p$-form $X$ on a ${\mathrm d}=2(p+1)$-dimensional worldvolume with split signature. For $p=0$, this describes a ``right-moving'' scalar on a $(1+1)$-dimensional worldsheet. For $p>0$ these models are higher-dimensional-worldvolume generalizations of the string. 
The stress-energy tensor on the worldvolume $\mathcal \T^{(+)}_{\un a \un b}= \tfrac1{4p!} F^{(+)}_{\un a}{}^{\un c_1\dots \un c_p}F^{(+)}_{\un b\un c_1\dots \un c_p}$ is constructed from the selfdual part of the $(p+1)$-form field strength $F=dX$. 
The Hamiltonian analysis of this system proceeds by singling out a time-like coordinate $\tau$ and splitting the worldvolume coordinates $(\sigma^{\un a}) \to (\tau, \sigma^a)$.
The stress-energy tensor splits as $\T^{(+)}_{\un a\un b} \to (\T_{ab}, \S^{a}, \T)$.

In previous work on U-duality-covariant strings \cite{Linch:2015fya, Linch:2015qva, Linch:2015fca}, we focused on the constraints $\S^a$ which simultaneously generate $\sigma$-translations on the worldvolume and dynamically impose the strong section condition in the target theory. 
In this section we supplement that analysis to include the Virasoro constraint $\T$ needed for a complete description of the F-brane dynamics.

%%%%%%%%%%%%%%%%%%%%%%%%%%%%%%%%%%%%%%%%%%%%%%%%%%%%%%%%%%%%%%%%
\subsection{Constraints}
\label{S:Constraints}
\begin{figure}[ht]
$$ \vcenter{\hbox{\includegraphics[width=2.5in]{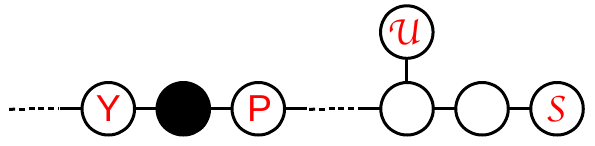}}} $$
\caption{\small Fundamental representations and currents: $P$ denotes the momentum current of the theory. $\S$ and $\U$ are the section and Gau\ss{} constraints. The scalars $Y$ will be added in section \ref{S:Critical}.}
\label{F:Representations}
\end{figure}

The chiral $p$-form interpretation is only straightforward for strings with ${\mathrm D}<5$. In general, the more appropriate language for F-theory is in terms of the fundamental representations $R_i$ of $E_n$ (for $i=1,\dots, n$ and $n={\mathrm D}+1$). 
In the numbering of the nodes of the $E_{11}$ diagram indicated in figure \ref{F:Fdiamond}, $\S^a$ is valued in the $R_1$ representation, and the dynamical parts of the gauge field $X^A$ (or its conjugate momentum $P_A$) are valued in the $R_n$ representation. 
We summarize the currents/constraints and the $E_n$ representations in which they are valued in figure \ref{F:Representations}. 

The momenta $P_A$ conjugate to $X^A$ are defined by the equal-$\tau$ commutator
 \begin{align}
[P_A(1) , X^B(2)] = - i \delta_A^B \delta(1-2)
.
\end{align}
In the Hamiltonian analysis, the selfdual part $F^{(+)}$ of the field strength defines the $E_n$-covariant current
\begin{align}
\label{E:LICurrent}
\P_A \equiv  P_A + \eta_{ABc} \partial^c X^B
,
\end{align}
while the anti-selfdual part 
\begin{align}
\label{E:ASD}
\widetilde \P_A \equiv  P_A - \eta_{ABc} \partial^c X^B
\end{align}
is the second-class constraint.
The $\eta$ symbols are the Clebsch-Gordan-Wigner coefficients mapping $R_n \otimes R_n \to  R_1$.
They are used to define the section constraints $\S^c \equiv \tfrac14  \P_A \eta^{ABc} \P_B$. 

To extend to $\T \equiv \S^0$, we need a symmetric $G$-invariant pairing $\eta^0$ on $R_n$. We will simply assume such a tensor exists (it does for low ranks) and derive the conditions it must satisfy.
In terms of these $G$-tensors, the section and Virasoro constraints are $\S^c \equiv \tfrac14  \P_A \eta^{ABc} \P_B$ and $\T \equiv \tfrac14 \P_A {\eta}^{AB 0} \P_B$.
To reduce clutter, we will often suppress $R_n$ indices, with tensors matrix-multiplied. 
Furthermore, we combine worldvolume indices $\un c= 0 ,c$ runing over both $\tau$ and $\sigma$ values so that we can write
\begin{align}
\label{E:Constraints}
\setlength{\fboxsep}{2.5\fboxsep} %add box padding
\boxed{
	\S^{\un c} =\tfrac14 \P \eta^{\un c} \P
}~.
\end{align}

The equal-$\tau$ commutator of the selfdual currents is
\begin{align}
[\P_A (1), \P_B(2)] = 2i  \eta_{AB c} \partial^c \delta(1-2) .
\end{align}
From this we compute the brackets of the currents with the constraints 
\begin{align}
[\S^{\un a} (1), \P(2) ] = i \eta_{c} \eta^{\un a} \partial^c \delta(1-2) \, \P(1)
\end{align}
On functions $f(X)$
\begin{align}
[\P_A , f ] &= - i \partial_A f  \, \delta 
,
\end{align}
so
\begin{align}
[\S^{\un a} , f ] = -\tfrac i2  (\partial f \eta^{\un a} \P) \, \delta
.
\end{align}
We can relate this generator to worldvolume diffeomorphisms by 
introducing its analog constructed from the second-class constraints (\ref{E:ASD}) and
observing that 
\begin{align}
\label{E:DeltaS}
\S^a - \tilde \S^a 
	= \tfrac14  (\P\eta^a\P - \widetilde \P\eta^a \widetilde \P) 
	= \tfrac14 (\P  + \widetilde \P)\eta^a(\P -\widetilde \P) 
=  \partial^b X \eta_b\eta^aP
	= \partial^a X P -   \partial^b X U^a_b P ,
\end{align}
where the last equality is the definition of $U$ (cf. ref. \cite{Linch:2015fca}).
It defines the Gau\ss{} law constraint of reference \cite{Linch:2015fya}:
\begin{align}
\label{E:U}
\setlength{\fboxsep}{2.5\fboxsep} %add box padding
\boxed{
\U^{a} \equiv U^{a}_{b} \partial^b P
~~~\textrm{with}~~~
U^{a}_{b} =\delta^{a}_b - \eta_{b}\eta^{a}
}~.
\end{align}
(This matrix satisfies $U^a_bU^b_c \propto U^a_c$, so it defines a projector $R_n \otimes R_1\to R_2$ associating $\U$ to the top node of figure \ref{F:Representations}: cf.\ \sec{}\ref{S:UOnly} and \sec{}\ref{S:UV}.)
Thus, we have found that
\begin{align}
\label{E:wvdiff}
[\S^a - \tilde \S^a, f ] &= -i \delta\, \partial^a f
\end{align}
up to Gau\ss{} law sectioning ({\it i.e.}\ $\U=0$ on the tensor algebra of the Hilbert space).

As we are working in the Hamiltonian formalism, the quantum brackets are equal-$\tau$ commutators and $\partial^0$ does not appear explicitly. Instead, $\tau$-evolution is determined by the Heisenberg operator
\begin{align}
\label{E:Heisenberg}
{d\over d\tau} = {\partial \over \partial \tau} + i H . 
\end{align}
Because of this, $\partial^0$ will never be generated on the right-hand-side of any equal-$\tau$ bracket. 
(We will define the Hamiltonian in terms of the constraints in \sec{}\ref{S:Action}, cf.\ eq.\ (\ref{E:Hamiltonian}).)

%%%%%%%%%%%%%%%%%%%%%%%%%%%%%%%%%%%%%%%%%%%%%%%%%%%%%%%%%%%%%%%%
\subsection{Algebra}
\label{S:Algebra}
We are now in a position to compute the equal-$\tau$ bracket of $\S^{\un a}$ with itself. 
Using the identity
\begin{align}
\partial^{\un a} \delta (1-2) A(1)B(2) 
	&= \partial^{\un a} \delta (1-2) AB
		+\tfrac 12 \delta (1-2)  A \on\leftrightarrow{\partial}{}^{\un a} B
\end{align}
with currents on the right-hand-side evaluated at the ``mid-point'' $\tfrac12[(1)+(2)]$,
we find
\begin{align}
\label{E:SScalc}
[\S^{\un a}(1) , \S^{\un b}(2)] 
	&= \tfrac i2 \partial^c \delta (1-2) \P(1) \eta^{\un a} \eta_c \eta^{\un b} \P(2) 
\cr
	&= \tfrac i2 \partial^c \delta (1-2) \P\eta^{\un a} \eta_c \eta^{\un b} \P 
		+\tfrac i4 \delta (1-2)  \P \eta^{\un a} \eta_c \eta^{\un b} \on\leftrightarrow{\partial}{}^c\P .
\end{align}
The first term on the right-hand side is symmetric. For this to be a Schwinger term, ({\it i.e.}\ close onto $\partial \delta \S$), there must be a tensor $k$ such that
\begin{align}
\label{E:Schwinger}
\tfrac12 \eta^{(\un a} \eta_c \eta^{\un b)}  = k^{\un a\un b}_{c\un d} \ \eta^{\un d} .
\end{align}
The second term in (\ref{E:SScalc}) is anti-symmetric. Expanding the last two $\eta$'s, 
\begin{align}
\P \eta^{\un a} \eta_c \eta^{\un b} {\on\leftrightarrow\partial}{}^c \P
	&= \P \eta^{\un a} (\delta_c^{\un b} - U_c^{\un b}) {\on\leftrightarrow\partial}{}^c\P
	= - \P \eta^{\un a} U_c^{\un b} {\on\leftrightarrow\partial}{}^c\P
	=  \partial^c( \P \eta^{\un a} U_c^{\un b}\P) 
	-\P \eta^{\un a} U_c^{\un b} \partial^c\P ,
\end{align}
with the $\delta$ term canceling by symmetry of $\eta$. 
(We extend the previous definition of $U_a^b$ in the obvious way to 
$U_a^{\un b} = \delta_a^{\un b} -\eta_a{}\eta^{\un b}$.)
The first term can be expanded as 
\begin{align}
\eta^{[\un a} U_c^{\un b]} = \eta^{[\un a} \delta_c^{\un b]} - \eta^{[\un a} \eta_c\eta^{\un b]}
\end{align}
but the $\eta^3$ term is anti-symmetric. 
For the second term, we use the definition of the Gau\ss{} law to write 
\begin{align}
\P \eta^{\un a} U_c^{\un b} \partial^c\P = \P \eta^{[\un a} \left ( \U^{\un b]} + \V^{\un b]} X\right) ,
\end{align}
where 
\begin{align}
\setlength{\fboxsep}{2.5\fboxsep} %add box padding
\boxed{
\V^{\un a} \equiv \tfrac12 V^{\un a}_{bc} \partial^b \partial^c
~~~\textrm{with}~~~
V^{\un a}_{bc} = U_{(b}^{\un a} \eta_{c)} 
}~.
\end{align}
Collecting results, we find the equal-$\tau$ bracket of the $\S$ constraints
\begin{align}
\label{E:Algebra}
\setlength{\fboxsep}{2.5\fboxsep} %add box padding
\boxed{
[\S^{\un a} ,\S^{\un b}] = i \partial^{c}\delta  k^{\un a\un b}_{c\un d}\S^{\un d} 
	-\tfrac i2 \delta \partial^{[\un a}\S^{\un b]}
	-\tfrac i4 \delta \P\eta^{[\un a} \U^{\un b]}
	-\tfrac i4 \delta \P\eta^{[\un a} \V^{\un b]}X
}~.	
\end{align}
This algebra holds modulo sectioning and with the understanding that $\partial^0$ is to be set to zero (cf.\ comment under eq.\ \ref{E:Heisenberg}). 

In section \ref{S:F-orms}, we will derive conditions on the Schwinger tensors $k$.
Their explicit form is dimension-dependent, but a common feature is that they satisfy 
\begin{align}
\label{E:ghg}
\eta^0 \eta_a \eta^0 = \eta_{ab} \eta^b
~~~\textrm{or}~~~
k^{00}_{a\un b} =  \eta_{a\un b} 
.
\end{align}
(This is a generalization of the ``factorization of the vielbein'' eq. 4.3 reference \cite{Linch:2015fya}.)
Here, $ú_{ab}$ is the flat metric on the worldvolume (which we write explicitly to distinguish it from $ú_{ABc}$ and $ú^{ABc}$).

To complete the analysis of the worldvolume constraint algebra, we need to know the quantum brackets with the Gau\ss{} law constraint and worldvolume sectioning. However,
\begin{align}
[\U^a, \P] = i  \V^a \delta
.
\end{align}
Thus, solving $\V^a=0$ guarantees that $\U^a$ commutes with the selfdual currents and therefore with everything. 
(We will give an alternative derivation of this constraint in \sec{}\ref{S:F-orms}.)
Thus, (\ref{E:Algebra}) gives the non-trivial part of the algebra of constraints defining the theory F. Truncating $\S^{\un a}\to \S^a$, we recover the algebra of references \cite{Linch:2015fya, Linch:2015qva, Linch:2015fca}.

Worldvolume reparameterizations by a vector field $(\xi_{\un a})= (\xi_0 , \xi_a)$ are generated by 
\begin{align}
\label{E:Reparam}
\delta_\xi = i \int d^{\rm d-1}§ \,\xi_{\un a} \S^{\un a}
.		
\end{align}
The worldvolume diffeomorphism algebra closes up to constraints as 
\begin{align}
\label{E:diff}
[\delta_\xi, \delta_{\xi'}] = \delta_{\xi''}
~~~\mathrm{with}~~~
\xi''_{\un a} = 
	- \tfrac12 k^{\un c \un d}_{b\un a} \xi_{\un c} {\on\leftrightarrow\partial}{}^b \xi'_{\un b} 
	- \tfrac12 \partial^b (\xi_{[b} \xi'_{\un a]})
.
\end{align}
The worldvolume gauge algebra includes, besides the reparameterizations above, the transformations generated by the Gau\ss{} constraint with parameter $\lambda_a^A$:
\begin{align}
\label{E:GaussGauge}
\delta_\lambda = i \int d^{\rm d-1}§ \, \lambda_a \U^a .
\end{align}
In the remainder of this section, we will use these constraints to compute the algebra of target space diffeomorphisms.

%%%%%%%%%%%%%%%%%%%%%%%%%%%%%%%%%%%%%%%%%%%%%%%%%%%%%%%%%%%%%%%%
\subsection{Brackets}
\label{S:Brackets}
The symmetries generated by the selfdual currents on the worldvolume of a string are generalizations of the diffeomorphisms of the target space.
On general grounds, then, the target space symmetries generated by the currents must include the diffeomorphism symmetries of the $E_{n(n)}$ exceptional field theory. 
To show this explicitly, we first compute the 
commutator
%``C-bracket'' \cite{Siegel:1993th, Siegel:1993bj} 
of two worldvolume currents $V_I^A\P_A(\sigma_i)$ with ${I=1,2}$:
\begin{align}
[V_1\P (1), V_2\P(2)] &=
	2i \partial^a\delta V_{1}\eta_a V_{2} 
	-i\delta V_{[1} \partial V_{2]}\P
	+i \delta V_{[1}\eta_a\partial^a V_{2]}
\end{align}
This can be simplified using the relation 
\begin{align}
\partial^a f 
	= \partial^a X\partial f 
	=(\eta_b \eta^a  + U^a_b) \partial^b X\partial f 
	= \tfrac12 \partial f \eta^a ( \P - \tilde \P)
	+\partial^b X U^a_b  \partial f 	.
\end{align}
Then, modulo second class constraints and Gau\ss{} law sectioning, 
\begin{align}
\label{C-bracket}
[V_1\P (1), V_2\P (2)] =
	2i \partial^a\delta V_{1} \eta_a V_{2}
	-i\delta\left[\delta^A_C \delta^B_D
			-\tfrac 12 \eta^{ABa} \eta_{CDa} 
	\right]
	V_{[1}^C \partial_A V_{2]}^D \P_B .
\end{align}
%This is the C-bracket of worldvolume currents. 
%The exceptional/generalized Lie derivative and Lie bracket follow from this by integrating one or both of these currents over the worldvolume respectively. 
The ``D- and C-brackets"
 \cite{Siegel:1993th, Siegel:1993bj} 
follow from this by integrating one or both of these currents over the worldvolume respectively. 
Such integrated currents 
\begin{align}
\label{E:vfs}
\Lambda_i = i \int d^{\rm d-1}§\, \lambda_i \P
\end{align}
are the F-theory analogs of vector fields in the target space. 

When both currents are integrated, we obtain the C-bracket
\begin{align}
\label{E:GeneralLie}
[\Lambda_1, \Lambda_2] = \Lambda_{12}
~~~\mathrm{with}~~~
\lambda_{12}^A = - \left( \delta^A_C \delta^B_D
		-\tfrac12 Y^{AB}{}_{CD}
		\right)
	\lambda_{[1}^C \on\leftrightarrow\partial_B \lambda_{2]}^D 
\end{align}
where
\begin{align}
\label{E:Y-projector}
Y^{AB}{}_{CD} =\eta^{ABa}\eta_{CDa}
\end{align}
is the projector from the symmetric product of $R_n$ with itself onto $R_1$.
This $Y$-tensor was defined in reference \cite{Berman:2012vc}.
This is the F-theory analog of the generalized Lie bracket:
When it is truncated to massless modes, we recover the Courant bracket of exceptional geometry \cite{Berman:2012vc}.
Thus, the higher-dimensional worldvolume F-theory is a generalization of the string worldsheet in which target space diffeomorphisms are enlarged to the $E_{n(n)}$ transformations of ``exceptional field theory'' \cite{Berman:2011pe, Coimbra:2011ky, Berman:2012vc, Park:2013gaj, Hohm:2013jma, Hohm:2013pua, Hohm:2013vpa, Hohm:2013uia, Godazgar:2014nqa, Hohm:2014fxa, Blair:2014zba, Musaev:2014lna, Abzalov:2015ega
}.
Further truncation of the exceptional coordinates reduces the exceptional field theory (with ``dual'' coordinates) to the $E_{n(n)}$ version of generalized geometry (without) \cite{Hull:2007zu, Berman:2010is}.

When only one of the two currents is integrated, we obtain the (asymmetric) D-bracket.
From (\ref{C-bracket})
%and stripping off the current, 
we find
\begin{align}
\label{E:D-bracket}
[\Lambda, V\P] = ¶_ÂV\P
~~~\mathrm{with}~~~
\delta_\lambda V^A
&=\lambda^B {\on\leftrightarrow{\vspace{0mm}\partial_B}} V^A
	+ Y^{AB}{}_{CD}
		(\partial_B \lambda^C) V^D 
\end{align}
By comparison, we see that the massless truncation of this D-bracket is the generalized Lie derivative (also known as the ``Dorfman bracket'' in the exceptional geometry literature), at least when equation (\ref{E:Y-projector}) is satisfied. 
This equation holds for $\mathrm D\leq 5$ and holds in D = 6 (corresp. to $E_7$) up to a new constraint called $\W$ in reference \cite{Linch:2015fca}. (We will mostly ignore $\W$ in this work.)

Backgrounds are introduced as usual:
\begin{align} 
\P_{A} = E_{A}{}^{M} \P_{M}.
\end{align}
Under spacetime gauge transformations (cf.\ eq.\ \ref{E:D-bracket}),
\begin{align}
\delta_\lambda E_{A}{}^{M} 
&=\lambda^{N} {\on\leftrightarrow{\vspace{0mm}\partial_{N}}} E_{A}{}^{M}
	+ Y^{{M} {N}}{}_{{P} {Q}}
		(\partial_{N}\lambda^{P}) E_{A}{}^{Q}
.		
\end{align}
The background satisfies an ``orthogonality constraint'' of the form \cite{Linch:2015fya}
\begin{align}
\label{E:orthogonality}
\eta^{AB c} E_{A}{}^{M} E_{B}{}^{N}= \eta^{MN p} e^c{}_p
\end{align}
which can be interpreted as fixing part of the worldvolume metric in terms of the spacetime metric. (It cannot be other way around, as the worldvolume fills only part of the spacetime.)

%%%%%%%%%%%%%%%%%%%%%%%%%%%%%%%%%%%%%%%%%%%%%%%%%%%%%%%%%%%%%%%%
%%%%%%%%%%%%%%%%%%%%%%%%%%%%%%%%%%%%%%%%%%%%%%%%%%%%%%%%%%%%%%%%
\section{Worldvolume F-orms}
\label{S:F-orms}

In reference \cite{Linch:2015lwa}, we introduced the notion of F-orms as an F-(super)gravity generalization of de Rham forms in the target space. (This was done explicitly only for the case of $E_{4(4)} = SL(5)$.) 
Such target space F-orms exist because of the $\S$ section condition. 
%Similarly, we can consider worldvolume sectioning $\V$ \cite{Linch:2015qva}, leading to the notion of worldvolume F-orms.
We now consider the worldvolume analog of this construction \cite{Linch:2015qva}. 

%%%%%%%%%%%%%%%%%%%%%%%%%%%%%%%%%%%%%%%%%%%%%%%%%%%%%%%%%%%%%%%%
\subsection{\texorpdfstring{$H$}{H} Field Strengths}
\label{S:HFS}
The constraints $\U$, $\V$, ... can be interpreted as the existence of a differential on the space of exceptional exterior forms, or worldvolume ``F-orms''.
We have seen this already in the structure of the currents:  The bosonic (anti)selfdual field strengths (currents) (\ref{E:LICurrent}) and (\ref{E:ASD}) in the $H$(amiltonian) form 
\begin{align}
F^{(à)} = P à \eta_{c} \partial^c X ­ \eta_0F_  à F_§
\end{align}
come with the gauge invariance generated by (\ref{E:U}): %, in matrix notation,
\begin{align*}
\U^a = U^{a}_{b} \partial^b P
~~~\textrm{with}~~~
U^{a}_{b} =\delta^a_b - \eta_{b}\eta^{a} .
\end{align*}
When translated to Lagrangian language (see below), $F_ $ will be the part of the full field strength carrying a $ $ (0) index, while $F_§$ is the part of the field strength defined in the $§$ subspace.  

We can thus construct a series of differential relations in this subspace (to be extended to the full worldvolume below) representing gauge transformations, field strengths, and Bianchi identities.  (This generalizes the special case of de Rham forms; see below.)
Using the notation
\begin{align} 
d \equiv ú_a »^a¼,â
d^a \equiv U^{Ta}_b \partial^b = »^a - ú^a d¼,â
d^{Ta} \equiv  \partial^b U^a_b = »^a - dú^a
\end{align}
we have
\begin{align} 
¶X = d^a Â_a¼,âF_§ = dX¼,âB^a = d^{Ta}F_§
\end{align}
defined on the gauge parameters $Â^A_a$, fields $X^A$, field strengths $F_{§A}$, and Bianchi identities $B_A^a$.
We thus have an analog of Hodge duality
\begin{align} 
\label{E:Hodge1}
Â_a^A ª B_A^a¼,âX^A ª F_{§A}
\end{align}
For these equations to be compatible, the $d$'s must satisfy the identities
\begin{align} 
\label{E:ddV}
dd^a = d^{Ta} d = \V^a  \to  0 
\end{align}
(the two expressions are equivalent), which requires the $\V$ constraint.

%%%%%%%%%%%%%%%%%%%%%%%%%%%%%%%%%%%%%%%%%%%%%%%%%%%%%%%%%%%%%%%%
\subsection{Only \texorpdfstring{$\U$}{U}}
\label{S:UOnly}

The simplest examples are those where the $\V$ and $\W$ constraints are absent.
For our purposes, that means D $²3$ for the $X$'s (or arbitrary scalars $Y$ and their duals $\widetilde Y$; we defer their discussion to section \ref{S:Critical}). 

Peeling off the two $»$'s from the $d^2$ identity (\ref{E:ddV}), we then have
\begin{align}
\label{E:Eta3}
¶^c_{(a}ú_{b)} - ú_{(a}ú^c ú_{b)} 
	= V^c_{ab}
	=0.
\end{align}
(Note that the second term is the (dual of the) Schwinger tensor eq.\ \ref{E:Schwinger}.)
From this identity follows
\begin{align} 
ú^c ú_a ú^b ú_c = (I - ú^c ú_c)ú^b ú_a + (ú^c ú_c)¶^b_a
\end{align}
so
\begin{align} 
U^2 = (I+ú^a ú_a)U 
\end{align}
If $ú^a ú_a ¾ I$, we also have
\begin{align}
ú^a ú_a = \f{\rm d}2 I 
\end{align}
so $U/(\f{\rm d}2+1)$ is a projection operator.

These cases are essentially just differential forms, so we now consider the latter directly.
In that case we replace the matrix product with the wedge product.  We next translate the standard representation of the algebra of differential forms as Clifford algebras into our notation.  We define
\begin{align}
ú_a = d§_a¼,âú^a = ¶^a
,
\end{align}
where $¶^a$ is the dual to $d§_a$:  It acts only on it, as
\begin{align}
¶^a d§_b = ¶^a_b - d§_b ¶^a
.
\end{align}
(The sign is from the usual antisymmetry of the wedge product.)
This is the anticommutation relation
\begin{align}
Ó ¶^a , d§_b Õ = ¶^a_b
\end{align}
of fermionic creation and annihilation operators, whose representation space is that of antisymmetric tensors of all ranks.  (It is also the direct sum of two Clifford algebras, representing left and right multiplication of those algebras on a representation space of square matrices, which can be expressed in a basis of products of Dirac $©$-matrices.  But the identification instead with fermionic oscillators manifests the $GL$ symmetry instead of just the $SO$ of Clifford algebras.)
As a result, the $ú^3$ identity (\ref{E:Eta3}) is easily satisfied, since
\begin{align}
d§_{(a}\wedge d§_{b)} = 0 .
\end{align}
We then find that 
\begin{align}
d ­ ú_a »^a = d§_a »^a
\end{align}
is the usual exterior derivative, as well as 
\begin{align}
d^a = dú^a¼,âd^{Ta} = ú^a d
\end{align}
For our purposes, the ``$AÊ$" indices on the $ú$'s would run over only selfdual and anti-selfdual tensors, or other dual pairs (like $Y$ and $\widetilde Y$), not the full range of all antisymmetric tensors.

%%%%%%%%%%%%%%%%%%%%%%%%%%%%%%%%%%%%%%%%%%%%%%%%%%%%%%%%%%%%%%%%
\subsection{\texorpdfstring{$\U$}{U} and \texorpdfstring{$\V$}{V}}
\label{S:UV}
Generalizing to include the $\V$ constraint will not modify the $d$ identities, but does change the $ú^3$ identity used to derive them:
Here we consider the cases with the $\V$ constraint but without $\W$ \cite{Linch:2015fca} ({\it i.e.}\ D = 4 and 5).

For the case D = 4, we have\footnote{Note that this is compatible with (\ref{E:ghg}) and can be combined with it to give 
\begin{align*}
ú_{(\un a}ú^c ú_{\un b)} - ¶^c_{(\un a}ú_{\un b)} = - ú_{\un a\un b}ú^{cd}ú_d .
\end{align*}
} %end footnote
\begin{align}
\label{E:eta3UV}
V^c_{ab} = \tfrac12 ú_{ab}ú^{cd}ú_d
.
\end{align}
The extra term produces the constraint \cite{Linch:2015qva}
\begin{align} 
\label{E:SomeV}
\V^a = \eta^{ab} \eta_b \V
~~~\textrm{with}~~~
\V \equiv üú_{ab}»^a »^b 
\end{align}  
when this identity is multiplied by $»^a »^b$: Thus
\begin{align} dd^a = d^{Ta} d = ú^{ab}ú_b\V £ 0 \end{align}

The new term breaks the $G$ symmetry to an orthogonal group.  Its coefficient is fixed by the requirement that the $U$ matrix be proportional to a projection operator, which also requires the $©$-matrix-like identity
\begin{align}
ú^a ú_b + (ú_{bc}ú^c)(ú^{ad}ú_d) = ¶_b^a .
\end{align}
(This implies the previous identity (\ref{E:eta3UV}).)
It also implies
\begin{align}
ú^a ú_a = \f{\rm d-1}2 I
\end{align}
\begin{align}
ú^c ú_a ú^b ú_c = - \f{\rm d-5}2 ú^b ú_a + \f{\rm d-3}2 ¶^b_a
\end{align}
so 
$U/\f{\rm d-1}2$ is a projection operator.

In D = 5 we have for the group $E_6$, instead of the orthogonal metric $ú_{ab}$, the totally symmetric symbol $d_{abc}$ for the {\bf 27} representation and $d^{abc}$ for the {\bf 27$'$}.  They satisfy the ``Springer relation" \cite{cvitanovi2008group, Springer1, Springer2}
\begin{align} 
\f18 d^{efg}d_{e(ab}d_{cd)f} = \f16 ¶^g_{(a}d_{bcd)} 
\end{align}
or, eliminating redundant terms,
\begin{align} 
d^{efg}(d_{eab}d_{cdf} + d_{eac}d_{dbf} + d_{ead}d_{bcf}) = ¶_a^g d_{bcd} + ¶_b^g d_{cda} + ¶_c^g d_{dab} + ¶_d^g d_{abc} .
\end{align}
In this case, since the $a$ and $A$ indices are the same, we identify $ú_{abc}=d_{abc}$, so this is our $ú^3$ identity:
\begin{align} 
V^c_{ab} = \tfrac12 ú_{abe}ú^{cde}ú_d - \tfrac12 I^c ° ú_{ab} 
,
\end{align}
where we have abbreviated
\begin{align} 
(I^c ° ú_{ab})_{de} ­ ¶^c_{(d}ú_{e)ab} 
.
\end{align}
Now the $\V$ constraint is the worldvolume dual of the $\S$ constraint, namely
\begin{align} 
\V_a  = üú_{abc}»^b »^câ
\leftrightarrow
â\S^a = \f14 ú^{abc}\P_b\P_c 
,
%\end{align}
%\begin{align} 
\\
dd^a = d^{Ta}d = ú^{abc}ú_b\V_c - I^a ° \V £ 0 
,
\end{align}
where we have used the shorthand notation
\begin{align} 
(I^a ° \V)_{bc} ­ ¶^a_{(b}\V_{c)} 
.
\end{align}

%%%%%%%%%%%%%%%%%%%%%%%%%%%%%%%%%%%%%%%%%%%%%%%%%%%%%%%%%%%%%%%%
\subsection{\texorpdfstring{$L$}{L} Field Strengths}
\label{S:LFS}
We now generalize the above results to the full worldvolume ($ $ and $§$), but still written in $H$-covariant notation.
The gauge parameters are $Â^A_a$; the bosons are now $X^A$, $X^A_a$ ($\U$ multipliers); their field strengths are $F_ ^A$, $F_{§A}$.
Then we have
\begin{subequations}
\begin{align}
¶X = »^a ( Â_a - ú^b ú_a Â_b )¼,â¶X_a = »^0 Â_a
\\
F_§ = »^a ( ú_a X )¼,âF_  = »^0 X - »^a ( X_a - ú^b ú_a X_b ) 
\\
( »^b ú_b ú^a - »^a ) F_§ = 0¼,â »^0 F_§ - »^a ( ú_a F_  ) = 0   
\end{align}
\end{subequations}
(Compare 4D electromagnetism in 3-vector notation.
Supersymmetrization will be considered below.)  

With the help of the $d$ notation, these equations simplify to
\begin{subequations}
\label{E:Complex}
\openup3\jot
\begin{align}
\delta 
\left(
	\begin{array}{c}
	X\\X_a
	\end{array}
\right)
	&=
\left(
	\begin{array}{c}
	d^a\\ \partial_0
	\end{array}
\right) \lambda_a
\\
%\end{align}
%%%%%%%
%\begin{align}
\left(
	\begin{array}{c}
	F_\sigma \\ F_\tau
	\end{array}
\right)
	&=
\left(
	\begin{array}{cc}
	d &0 \\ \partial_0 &-d^a
	\end{array}
\right) 
\left(
	\begin{array}{c}
	X \\ X_a
	\end{array}
\right)
\\
%\end{align}
%%%%%%%
%\begin{align}
\left(
	\begin{array}{c}
	B \\ B^a
	\end{array}
\right)
	&=
\left(
	\begin{array}{cc}
	-\partial_0 & d \\ d^{Ta} & 0
	\end{array}
\right) 
\left(
	\begin{array}{c}
	F_\sigma \\ F_\tau
	\end{array}
\right)
\\
%\end{align}
%%%%%%%
%\begin{align}
	{BB}^a
	&=
\left(
	\begin{array}{cc}
	d^{Ta} & \partial_0
	\end{array}
\right) 
\left(
	\begin{array}{c}
	B \\ B^a
	\end{array}
\right)
\end{align}
\end{subequations}
where the $B$'s are the Bianchi identities, and ${BB}$ is the Bianchi identity of the Bianchi identities.
This implies a generalization of the Hodge duality (\ref{E:Hodge1}) considered in the $§$ space, as expected from adding 1 dimension:
\begin{align}
\lambda^A_a 
	\leftrightarrow {BB}^a_A
~,~~
\left(
	\begin{array}{c}
	X^A \\ X^A_a
	\end{array}
\right)
	\leftrightarrow 
\left(
	\begin{array}{c}
	B_A \\ B_A^a
	\end{array}
\right)
~,~~
\left(
	\begin{array}{c}
	F_{\sigma \, A} \\ F^A_\tau
	\end{array}
\right)
	\leftrightarrow 
\left(
	\begin{array}{c}
	 F^A_\tau \\ -F_{\sigma \, A}
	\end{array}
\right) .
\end{align}

In the special case of true differential forms given above, this is easily seen to reduce to the usual (in $§¢ $ notation):  Then we have the replacments
\begin{align}
\lambda_a \to \lambda = \eta^a \lambda_a 
~,~~
X_a \to X_0 = \eta^a X_a
~,~~
B^a = \eta^a B^0
~,~~
{BB}^a = \eta^a {BB}. 
\end{align}
The $§$ forms can then be upgraded to $§ $ forms by
\begin{align}
öd = d + d Ê»_0
\end{align}
and
\begin{align}
ßX = X + d \^ X_0¼,âF = F_§ + d \^ F_ ¼,âßB = B^0 - d \^ B .
\end{align}
Then the equations (\ref{E:Complex}) collapse to
\begin{align}
¶ßX = öd¼,âF = ödßX¼,âßB = ödF¼,âödßB = d \^ BB + dB^0 .
\end{align}

%%%%%%%%%%%%%%%%%%%%%%%%%%%%%%%%%%%%%%%%%%%%%%%%%%%%%%%%%%%%%%%%
%%%%%%%%%%%%%%%%%%%%%%%%%%%%%%%%%%%%%%%%%%%%%%%%%%%%%%%%%%%%%%%%
\section{Critical Superstrings}
\label{S:Critical}

Critical U-duality-covariant superstrings were constructed for ranks $n\leq 7$ in \cite{Linch:2015fca}.
They are obtained from the non-critical D-dimensional U-duality-covariant bosonic superstrings by completing with 
${\mathrm D}' = 10- \mathrm D$ spacetime coordinate fields $Y^{a'}$ (and their duals $\widetilde Y_{aa'}$)
and 32 fermionic Green-Schwarz fields $\Theta^{\un \alpha}$. 
We provide a summary of indices and symbols in appendix \ref{S:Notation}.

%%%%%%%%%%%%%%%%%%%%%%%%%%%%%%%%%%%%%%%%%%%%%%%%%%%%%%%%%%%%%%%%
\subsection{Supersymmetry}
\label{S:kappa}
Finally, we introduce the 32 Green-Schwarz fermionic worldvolume scalars $\Theta^{\un \alpha}$. These can be contracted with (spacetime) Pauli-like matrices $(\gamma^{\un A})_{\un \alpha\un \beta}$. 
Introducing the dual matrices $(\tilde \gamma^{\un A})^{\un \alpha\un \beta}$, they close onto worldvolume Dirac-like matrices $(\Gamma_{\un a})_{\un \alpha}{}^{\un \beta}$ as \cite{Linch:2015fca}
\begin{align}
\gamma^{\un A} \tilde \gamma^{\un B}
+\gamma^{\un B} \tilde \gamma^{\un A}
= 2 \eta^{\un A\un B \un c} \Gamma_{\un c} ,
\end{align}
where we take the $\tau$ component $(\Gamma_0)_{\un \alpha}{}^{\un \beta} = \delta_{\un \alpha}^{\un \beta}$ to be the unit matrix.

The supersymmetry currents are
\begin{subequations}
\label{E:D}
\begin{align}
D & = \Pi  +\gamma \Theta \, (\P -  \eta_{c} \chi^c) \\
\P & = P + \eta_a( \partial^a X + 2\chi^a) \\
\Omega^{a\un \alpha}& =-2i \partial^a \Theta^{\un \alpha} \\
÷\P & = P - \eta_a\partial^a X 
\end{align} 
\end{subequations}
where
\begin{align}
\chi^{\un A \un a} \equiv - i \Theta \gamma^{\un A} \partial^{\un a} \Theta
.
\end{align} 
The ``selfdual" supercurrents $DP¯$ satisfy the deceivingly ``familiar'' bracket relations \cite{Linch:2015fca}
\begin{align}
\{ D , D \} = 2 \, \gamma \, \P \, \delta
~~~\textrm{and}~~~
[D , \P ] = 2\, \eta_{c} \, \gamma \, \Omega^{c} \, \delta
,
\end{align}
while the anti-selfdual (non-super) current $÷\P$ commutes with the rest.

The Hamiltonian action is invariant under the supersymmetry transformation generated by
%\begin{align}
%\label{E:q}
%q = \int \left\{ 	
%	\Pi - (\Theta \gamma) \left[ 
%		P + \eta_{a}
%			\left(
%				\partial^a X +\tfrac 13 \chi^a 
%			\right)
%		\right]
%	\right\}
%\end{align}
\begin{align}
\label{E:q}
q = \int \left[ 	
	\Pi - \gamma \Theta \left( 
		\P  -\tfrac 53 \eta_a\chi^a 
		\right)
	\right]
\end{align}
inducing
\begin{align}
\delta \Theta = \epsilon
~,~~
\delta X^{\un A} = i \epsilon \gamma^{\un A} \Theta 
~\textrm{, and}~~
\delta P_{\un A} = -i \eta_{\un A\un B c}\, \epsilon\gamma^{\un B} \partial^c \Theta
.
\end{align}

As usual, the constraints $D=0$ are mixed first and second class.
The first class part defines the $\kappa$ symmetry generator 
\begin{align}
\mathcal B = \sla \P D 
	= \P_{\un A} \tilde \gamma^{\un A} D .
\end{align}
Its components close onto the $\S$ generators
\begin{align}
\{ \mathcal B^{\un \alpha}, \mathcal B^{\un \beta} \} =  
	2( \sla \P \Gamma_{\un a})^{(\un \alpha \un \beta)} {\,\,} \S^{\un a} {\,\,} \delta +\dots
\end{align}
modulo terms containing the second class constraint $D$. 
(The $\mathcal B$ part of the algebra is formally the same as the analogous string algebra in reference \cite{Siegel:1985xj} under the replacement $\mathcal A\to \mathcal S$.)
The worldvolume gauge algebra includes, besides the reparameterizations (\ref{E:Reparam}) and the transformations (\ref{E:GaussGauge}) generated by the Gau\ss{} law constraint, the $\kappa$ symmetry transformations
\begin{align}
\delta_\kappa = i \int d^{\rm d-1}\sigma \, \kappa_{\un \alpha} \B^{\un \alpha} . 
\end{align}

%%%%%%%%%%%%%%%%%%%%%%%%%%%%%%%%%%%%%%%%%%%%%%%%%%%%%%%%%%%%%%%%
%%%%%%%%%%%%%%%%%%%%%%%%%%%%%%%%%%%%%%%%%%%%%%%%%%%%%%%%%%%%%%%%
\subsection{\texorpdfstring{$H$}{H} Action}
\label{S:Action}
With the constraints $\S^{\un a}$ closing on Gau\ss{}'s law $\U^a$ and worldvolume sectioning $\V^a$, we are in a position to write down the action of the worldvolume theory in Hamiltonian form. To this end, we first impose worldvolume sectioning \cite{Linch:2015qva}
\begin{align}
\V^a = 0
\end{align}
and define the action on the space of solutions to this constraint.
Then, in Hamiltonian form it is given by $S_H = \int L_H$ with 
\begin{align}
L_H = 
	-\dt X P
	+i \dt \Theta \Pi 
	+H
\end{align}
in terms of the Hamiltonian
\begin{align}
\label{E:Hamiltonian}
H = \lambda D
 	- X_a \U^a 
	+ \tfrac12  \ell_{0} ( \T +\tilde \T )
	+ \tfrac12  \ell_{a} ( \S^a - \tilde \S^a )
.
\end{align}
Here $X_a$, $\ell_0$, and $\ell_a$ are the Lagrange multipliers for the constraints
$\U$ (\ref{E:U}) and $\S$, $\T$ (\ref{E:Constraints}), respectively. 
(This name for the $\U$ multiplier will be justified in \sec{}\ref{S:LAction}.)
We recall here that the tilded versions of the constraints are defined by replacing the selfdual currents $\P$ with the anti-selfdual currents $\tilde \P$.
This Lagrangian is manifestly spacetime supersymmetric and $H$-invariant. $G$-invariance requires coupling to a background as introduced in section \ref{S:Brackets}.

%%%%%%%%%%%%%%%%%%%%%%%%%%%%%%%%%%%%%%%%%%%%%%%%%%%%%%%%%%%%%%%%
\subsection{\texorpdfstring{$L$}{L} Action}
\label{S:LAction}
We now set $D\to 0$ as a second-class constraint to solve for the momentum $¸$ conjugate to $\Theta$, and integrate out $P$
using its equation of motion.
Since $\P$ and $÷\P$ are to be identified with the selfdual and anti-selfdual field strengths, we write their resulting forms as
\begin{align}
\label{E:ASDcurrents}
\P = ú_0 F_  + F_§¼
~~~\textrm{and}~~~
÷\P = ú_0 F_  - F_§ .
\end{align}
We then find that
\begin{align}
\label{E:FSs}
F_{\tau} = \partial^0 X +\chi^0 -\partial^a( X_a - \eta^b \eta_a X_b) 
~~~\textrm{and}~~~
F_{\sigma} = \eta_a(\partial^a X +\chi^a)
\end{align}
(as well as
$P = ú_0 F_  - ú_a ^a$)
are simply the result of supersymmetrizing the Lagrangian bosonic field strengths obtained previously by the substitution $»X£»X+$, as expected from the usual invariant worldvolume currents $dX -i \Theta \gamma d \Theta$ and $d\Theta$ as the left-invariant 1-forms on the target space.

The Lagrangian is then
\begin{subequations}
\begin{align}
L & = \ell_0 ü F_§ ú^0 F_§ - \ell_0^{-1}ü(F_  - \ell_m ú^m F_§)ú_0(F_  - \ell_n ú^n F_§) + L_{WZ} 
\\
L_{WZ} & = -\on\circ F_§ ^0 + \on\circ F_  ú_a ^a  = -\on\circ F_§ F_  + \on\circ F_  F_§ ,
\end{align}
\end{subequations}
where in the Wess-Zumino term $L_{WZ}$, $\on\circ F$ stands for the $\chi = 0$ part of $F$.

This action can be rewritten in the form
\begin{align}
\label{E:LTLagrangian}
L = å{-g}Êü(ßF_§ ú^0 ßF_§ - ßF_  ú_0 ßF_ ) +L_{WZ}
\end{align}
where
\begin{align}
å{-g} = \ell_0¼,âßF_§ = F_§¼,âßF_  =  \ell_0^{-1}(F_  - \ell_m ú^m F_§)  ,
\end{align}
from which we find that the worldvolume vielbein $e^{\un a}{}_{\un m}$ takes the simple form
\begin{align}
\label{E:wvFrame}
e^a{}_{\un m}=  ¶^a{}_{\un m}¼,âe^0{}_{\un m} = \ell_0^{-1}(1,-\ell_m)
\end{align}
(the former from $ßF_§$, the latter from $ßF_ $).

For example, in the case of standard differential forms considered previously (cf.\ section \ref{S:UOnly}), flattening the indices on $ßF_§$ (none of which are ``0") is trivial, while on $ßF_ $ (only one ``0" index), the $ú^m$ ($¶^m$) picks off one ``$m$" index from $F_§$.

Note that there are no $Î^4$ terms in the Wess-Zumino term, since this is a ``heterotic" type of construction.  They will reappear upon reduction of F-theory to T-theory below:  As usual in dimensional reduction, selfdual theories reduce to non-selfdual theories.

%%%%%%%%%%%%%%%%%%%%%%%%%%%%%%%%%%%%%%%%%%%%%%%%%%%%%%%%%%%%%%%%
%%%%%%%%%%%%%%%%%%%%%%%%%%%%%%%%%%%%%%%%%%%%%%%%%%%%%%%%%%%%%%%%
\subsection{Actions from Selfduality}
\label{S:ActionsSD}
To gain some insight into the structure of the Lagrangian (\ref{E:LTLagrangian}), we derive it covariantly. 
A useful method to treat actions for selfdual theories is to use covariant, non-selfdual actions, and then impose selfduality separately.
A well-known way to derive the corresponding field equations compatible with selfduality is by replacing field strengths with their duals in the Bianchi identities.
We thus begin with the general gauge transformations, field strengths, and Bianchi identities for the bosons in F-theory, supersymmetrized, in $H$-covariant form.
For simplicity, we work in the ``conformal" gauge $e^{\un a}{}_{\un m} = \delta^{\un a}{}_{\un m}$ for the worldvolume metric.

After reintroducing $Î$, the gauge transformations are unchanged ($¶Î=0$), and the field strengths are supersymmetrized by replacing $»X£»X+$ (cf.\ eq.\ \ref{E:FSs}). For ease of reference, we reproduce them here in the new notation:
\begin{align}
\label{E:FSs2}
F_{\sigma} = dX +\eta_a\chi^a
~~~,~~~
F_{\tau} = \partial^0 X -d^aX_a +\chi^0.
\end{align}
The Bianchi identities then become the analog of $dF=d$:
\begin{align}
( »^b ú_b ú^a - »^a ) F_§ = ü( ú_b ú^a ú_c - ¶_a^b ú_c ) »^{[b} ^{c]}
~~~\textrm{and}~~~
»^0 F_§ - »^a ( ú_a F_  ) = ú_a »^{[0}^{a]} ,
\end{align}
where
\begin{align}
»^{[\un a}^{\un b]} = -i(»^{[\un a}Î)©(»^{\un b]}Î)
\end{align}
showing supersymmetry invariance.

For consistency with the selfduality condition $÷\P=0$, that is
\begin{align}
F_§ = ú_0 F_ 
\end{align}
(cf.\ eq.\ \ref{E:ASDcurrents}), variation of the metric part of the non-selfdual action must give exactly the result of switching $F_§ªú_0 F_ $ in the Bianchi identities, less the $$ correction terms from the Wess-Zumino part of the action:  After integration by parts,
\begin{align}
¶ü(F_§ ú^0 F_§ - F_  ú_0 F_ ) £ ¶X [ »^0 ú_0 F_  - »^a ( ú_a ú^0 F_§ ) ] + ¶X_a ( »^b ú_b ú^a - »^a ) ú_0 F_  .
\end{align}
This implies the necessity of the Wess-Zumino terms
\begin{align}
L_{WZ} &= -X ú_a »^{[0}^{a]} - X_a ü( ú_b ú^a ú_c - ¶_b^a ú_c ) »^{[b} ^{c]} 
\cr&
	\to \on\circ F_  F_§ -\on\circ F_§ F_  
\end{align}
after integration by parts, showing gauge invariance. (Modifying the WZ terms by integration by parts will change the expressions for the currents (\ref{E:D}) and (\ref{E:q}) by a canonical transformation \cite{Linch:2015fca}.) 
The result for $L$ then agrees with that obtained by Legendre transformation (\ref{E:LTLagrangian}) in the gauge $\ell_0=1$, $\ell_m=0$ (the flat case).

%%%%%%%%%%%%%%%%%%%%%%%%%%%%%%%%%%%%%%%%%%%%%%%%%%%%%%%%%%%%%%%%
%%%%%%%%%%%%%%%%%%%%%%%%%%%%%%%%%%%%%%%%%%%%%%%%%%%%%%%%%%%%%%%%
\subsection{Reduction F \texorpdfstring{$\to$}{\textrightarrow} T}
\label{S:F2T}
The Lagrangian (\ref{E:LTLagrangian}) has a heterotic-type Wess-Zumino term due to the doubled nature of the Green-Schwarz fermions. In this section, we explain how the type-II action of the underlying string is generated. 

Dimensional reduction from F-theory to T-theory comes from solving the Gau\ss{} constraint $\U$.  (Also the $\V,\W$ constraints are solved, but we'll ignore those here for simplicity.)  Since both $»$ and $P$ are bispinors, we can write this constraint in matrix notation as
\begin{align}
\U = »P - P» = 0
.
\end{align}
(See appendix \ref{S:SummarySymmetry}.)

The doubled spinor index is thereby divided into left and right halves:  Choosing a solution where $»$ picks one particular direction, its single $©$-matrix can be chosen block diagonal, forcing $P$ to also be so.  Thus we have a single $§$ for the worldsheet, while $P$ has lost its Ramond-Ramond (LR) pieces, as seen from the commutation relations $ÓD,DÕ¾P$.

While the effect on the metric terms in the action is to simply drop the LR terms, the WZ term generates a $Î^4$ term:  Since $P¾»X+$, the $»X$ ($\on\circ F$) factor there is replaced by $-$ for the LR piece.  This results in $_{LR}\wedge _{LR}¾_{LL}\wedge _{RR}$, as in the usual Green-Schwarz action.

%%%%%%%%%%%%%%%%%%%%%%%%%%%%%%%%%%%%%%%%%%%%%%%%%%%%%%%%%%%%%%%%
%%%%%%%%%%%%%%%%%%%%%%%%%%%%%%%%%%%%%%%%%%%%%%%%%%%%%%%%%%%%%%%%
\section{Conclusions}
\label{S:Conclusion}
In this work, we have given the Lagrangian formulation for the F-theory of critical superstrings in which the $E_{n(n)}$ U-duality symmetry is manifest for rank $n\leq 7$. 
(This corresponds to critical superstrings in a $\mathrm D + (10-\mathrm D)$-dimensional split with $\mathrm D = n-1$.)
To do this, we first extended the Hamiltonian description of the current algebra of constraints generating worldvolume translations \cite{Linch:2015fca} to include the generator responsible for the dynamics of the fundamental brane. We then constructed the Hamiltonian action from the constraints and performed a Legendre transformation to the Lagrangian form. 
This gives the worldvolume metric in terms of the Lagrange multipliers of the Hamiltonian formulation (\ref{E:wvFrame}). 
The resulting theory has the peculiar property that only d components of the d-dimensional worldvolume metric are needed for a covariant description instead of the usual $\tfrac12 \mathrm d(\mathrm d+1)$. 
(Thus, covariance is not manifest on the worldvolume except perhaps in a manner analogous to \cite{Sen:2015nph}.)
Consequently, they can all be gauged away in an analog of conformal gauge for the string (in contrast to the usual attempts to quantize the membrane in which the worldvolume gravity is dynamical).

In this new formulation, the worldvolume fields can be interpreted as an F-theory generalization of the ordinary $p$-form gauge fields of Maxwell-like theories. 
The Gau\ss{} law constraint $\U$ is used to define the analog of the exterior derivative operator and the Laplace constraint $\V$ implies that it is a differential. 
This gives rise to the exceptional geometry analog of the usual de Rham forms on the worldvolume.

The Lagrangian of the supersymmetric theory resembles that of a heterotic superstring in that the Wess-Zumino term has no $\Theta^4$ terms. 
This is the structure determined by the consistency between supersymmetry and the selfduality of the field strength of the fundamental gauge field. 
Reducing from F to the string by solving the Gau\ss{} law sectioning condition, 
%In reducing F $\to$ T (the T-duality-covariant description of the critical type-II string \cite{Siegel:1993th, Siegel:1993bj}) by Gau\ss{} law sectioning, 
the familiar Wess-Zumino term of the type-II Green-Schwarz action is recovered.

All formulations of all these theories have a symmetry $H$ dependent only on the dimension D.
(There is also the symmetry $H'=SO(10- \mathrm D)$ which we leave implicit here.)
If we limit our discussion to just the bosonic sector of a theory, then this symmetry is extended to the group $G$ in the Hamiltonian formalism, and to $L$ in the Lagrangian formalism.
Thus $ú_{ABa}$ is an invariant tensor of $G$, while $ú_{AB\un a}$ is an invariant tensor of $L$.
However, after extending to supersymmetry by including the fermions, only the $H$ subgroup survives in either formalism:
While the bosons form representations of the larger groups, the fermions are representations of only $H$.
As familiar from extended supergravity, the larger symmetry can be restored by including fields of the corresponding coset:
In our case, the spacetime (second-quantized) vielbein $E$ lives on the coset $G/H$, while the worldvolume (first-quantized) vielbein $e$ lives on the coset $L/H$.
Thus the background fields $E$ restore $G$ symmetry to the Hamiltonian formalism (and the resulting spacetime theory), while the Lagrange multipliers $e$ restore $L$ symmetry to the Lagrangian formalism.
In particular, if we ignore all $GL(1)$'s in the definition of $L$, we find $L/H$ has $\mathrm d-1$ generators for $\mathrm D ²5$.
The consistency of the simultaneous $G/H$ and $L/H$ cosets is then enforced by the orthogonality constraint (\ref{E:orthogonality}).

\section*{Acknowledgements}
It is a pleasure to thank Machiko Hatsuda for discussions relating our F-theory work to the brane current algebra approach of references \cite{Hatsuda:2012uk, Hatsuda:2012vm, Hatsuda:2013dya},
and Andy Royston and Stephen Randall for discussions and suggestions that helped to improve the presentation.
W{\sc dl}3 is grateful to the Simons Center for Geometry and Physics for hospitality during the {\sc viii} and {\sc ix} Simons Summer Workshops where parts of this project were completed.
W{\sc dl}3 is supported by National Science Foundation grants PHY-1214333 and PHY-1521099.
W{\sc s} was supported by NSF grant PHY-1316617.

\appendix
%%%%%%%%%%%%%%%%%%%%%%%%%%%%%%%%%%%%%%%%%%%%%%%%%%%%%%%%%%%%%%%%
%%%%%%%%%%%%%%%%%%%%%%%%%%%%%%%%%%%%%%%%%%%%%%%%%%%%%%%%%%%%%%%%
\section{Notation}
\label{S:Notation}
The theories we describe in this paper are complicated by the various symmetries and fields involved. 
Firstly, there is the worldvolume physics in its $H$(amiltonian) and $L$(agrangian) descriptions. 
Worldvolume fields come in two basic parts: $X$, which are selfdual in the $L$ form, and $Y$ which are scalars, but selfduality requires we introduce their duals $\widetilde Y$ as well.
Additionally, there are fermionic coordinates $\Theta$ required for supersymmetry. 
The $X$ fields are also valued in a target space with its $E_{n(n)}$ symmetry.
As we need definitions of fields, symmetries, and their indices, we include for reference this appendix collecting all the notation.  

\begin{table}[ht]
$$ \vcenter{
\halign{ #\hfilâ & $#$\hfilâ & $#$\hfilâ & $#$\hfil \cr
\bf representation & \hbox{{\bf H} ($'$ for {\bf H}$'$)} & \hbox{\bf H$°$H$'$} & \hbox{\bf L} \cr
spinor & ŒÊÊ(orÊÊÀŒ) & \un Œ = ŒŒ' & ŒÊÊ(orÊÊÀŒ,Ќ,À{Ќ}) \cr
vector: $§$ & a & & \un a \cr
\phantom{vector:} $X$(,$Y,\widetilde Y$) & A & \un A & \cr
super & & \A = (\un Œ,\un A,\un Œ a) & \cr
}
}
$$
\caption{\small Our indices are Greek for spinor, Latin for vector (lower case for worldvolume, upper case for spacetime), and calligraphic for super. The $H$ $X$-vector index $A$ also labels $L$ (anti-)selfdual worldvolume tensors ({\it e.g.\ }the field strength $F$). The primed version denotes the $(Y,\widetilde Y)$ indices.}
\label{T:Indices}
\end{table}
%\vskip.2in

Spacetime coordinates and conjugate momenta are represented by $(X^{\un A},P_{\un A})$ and $(Î^{\un Œ},¸_{\un Œ})$. The $E_{11}$ split is reflected on the bosonic coordinates as $X^{\un A} = (X^A;Y^{A'}) = (X^A;Y^{a'},\widetilde Y_{aa'})$. 
We do not need the split of the fermionic coordinates in the main body of this paper, but as it is important to the symmetry structure of the theory, we refer to it in a condensed form in appendix \ref{S:SummarySymmetry}.
(It is worked out in detail in ref. \cite{Linch:2015fca}.)
Also in this paper, we do not differentiate between the momenta for $Y$ and $\widetilde Y$ but when it is useful, we call the former $\Upsilon$ and the latter $\digamma$.
The entire collection of covariant currents is lumped into the ``nacho'' $\dd_\A = (D_{\un Œ},\P_{\un A},¯^{\un Œa})$. (Again, we have little recourse to this symbol in this paper but we include it for completeness.) The definition of these currents is given in (\ref{E:D}).

In the $H$ form, because of the $E_{11}$ split, there is the $E_{n(n)}$ part $\mathbf H$ and the rest $\mathbf H'$. We collect the relevant notation for the indices of this separation in table \ref{T:Indices}.

%%%%%%%%%%%%%%%%%%%%%%%%%%%%%%%%%%%%%%%%%%%%%%%%%%%%%%%%%%%%%%%%
\section{Symmetry}
\label{S:SummarySymmetry}

\font\qc=manfnt scaled\magstep2

\catcode`\ =9 \endlinechar=-1 % ignore all spaces (temporarily)
\newcount\dir \newdimen\yy \newdimen\ww
\newif\ifvisible \let\b=\visibletrue \let\w=\visiblefalse
\newbox\NE \newbox\NW \newbox\SE \newbox\SW \newbox\NS \newbox\EW
\setbox\SW=\hbox{\qc a} \setbox\NW=\hbox{\qc b}
\setbox\NE=\hbox{\qc c} \setbox\SE=\hbox{\qc d}
\ww=\wd\SW \dimen0=\fontdimen8\qc
\setbox\EW=\hbox{\kern-\dp\SW \vrule height\dimen0 width\wd\SW} \wd\EW=\ww
\setbox\NS=\hbox{\vrule height\ht\SW depth\dp\SW width\dimen0}  \wd\NS=\ww
\def\l{\ifcase\dir \dy+\NW \or\dx-\SW \or\dy-\SE \or\dx+\NE\dd-4\fi \dd+1}
\def\s{\ifcase\dir \dx+\EW \or \dy+\NS \or \dx-\EW \or \dy-\NS \fi}
\def\r{\ifcase\dir \dy-\SW\dd+4 \or\dx+\SE \or\dy+\NE \or\dx-\NW\fi \dd-1}
\def\t{\ifcase\dir\kern-\ww\dd+2\or\ey-\dd+2\or\kern\ww\dd-2\or\ey+\dd-2\fi}
\edef\dd#1#2{\global\advance\dir#1#2\space}
\def\dx#1#2{\ifvisible\raise\yy\copy#2 \if#1-\kern-2\ww\fi\else\kern#1\ww\fi}
\def\dy#1#2{\ifvisible\raise\yy\copy#2 \kern-\ww \fi \global\advance\yy#1\ww}
\def\ey#1{\global\advance\yy#1\ww}
\def\path#1{\hbox{\b \dir=0 \yy=0pt #1}}
\catcode`\ =10 \endlinechar='15 % resume normal spacing conventions

%%%%%%%%%%%%%%%%

In this section, we give a condensed summary of the target and worldvolume symmetry structures of the $E_{\mathrm D+1}$ theories. 
The target space symmetry group is called $G$ ($=E_{\mathrm D+1}$ throughout this work). 
It is the symmetry group of the F-theory currents and their bracket algebra. 
The Hamiltonian representation of the currents preserves all of $G$ for the bosonic part but only a subgroup $H$ if the Green-Schwarz fermions are included. 
This subgroup can be interpreted as the ``rotation'' subgroup of a worldvolume Lorentz group $L$.
In this appendix, we review the relationships between the groups $G\supset H \subset L$ for various D.
Additional details can be found in reference \cite{Linch:2015fca}.

Because of the supersymmetric current algebra $ÓD,DÕ$, both the worldvolume and the (bosonic) spacetime coordinates $§$ and $X$ can be written as bispinors, where for (at least) $\mathrm D ² 7$ only reality and symmetry constraints need be applied.  These are also sufficient to define the $H$ group.  
\begin{figure}[ht]
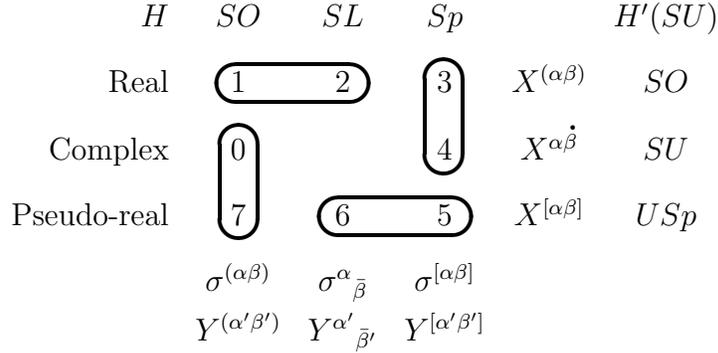

\begin{align*}
\kern-7em
\begin{matrix} \hfill H & SO & SL & Sp & & H' (SU) \cr
		\noalign{\vskip.5em}
		\hfill\textrm{Real} & 1 & 2 & 3 & X^{(μ)} & SO \cr
		\noalign{\vskip.5em}
		\hfill\textrm{Complex} & 0 & & 4 & X^{ŒÀº} & SU \cr
		\noalign{\vskip.5em}
		\hfill\textrm{Pseudo-real} & 7 & 6 & 5 & X^{[μ]}&  USp \cr
		\noalign{\vskip.5em}
		& §^{(Œº)} & §^Œ{}_{к} & §^{[Œº]} & & \cr
		& Y^{(Œ'º')} & Y^{Œ'}{}_{к'} & Y^{[Œ'º']} & & \cr
		\end{matrix}
%\kern-10.8em\raise1.8em\path{\s\s\s\l\l\s\s\s\l\l}
%\kern6.6em\raise-.8em\path{\l\s\s\l\l\s\s\l}
%\kern-3.6em\raise-2.8em\path{\s\s\s\l\l\s\s\s\l\l}
%\kern-2.5em\raise-3em\path{\l\s\s\l\l\s\s\l}\kern9em
\kern-15em\raise2.5em\path{\s\s\s\l\l\s\s\s\l\l}
\kern6.7em\raise.2em\path{\l\s\s\l\l\s\s\l}
\kern-3.4em\raise-1.8em\path{\s\s\s\l\l\s\s\s\l\l}
\kern-3.2em\raise-1.9em\path{\l\s\s\l\l\s\s\l}\kern9em
\end{align*}
\caption{\small Index structure of the coordinates of the theory organized by dimension D mod 8 (indicated by ovals).}
\label{F:dense}
\end{figure}
The diagram in figure \ref{F:dense} indicates their index structure and the corresponding $H$ groups for the physical spacetime dimension D mod 8.  The ovals indicate the range of spinor indices, enclosing the dimensions (D) with equal ranges, before considering reality properties; outside it increases by a factor of 2 when increasing D by 2.  The range is double that of the corresponding Lorentz group, so starts with 2 for $\mathrm D=1,2$.

The extension of $H$ symmetry to the $L$ symmetry of the Lagrangian drops the $SO$ and $Sp$ conditions, and gives two copies of $SL$ (except for $\mathrm D=10$).  Thus, in general the number of generators roughly doubles.
The $§$'s above include $ $, which appears as the trace piece upon introducing the $SO$ or $Sp$ metric, or reducing the 2 $SL$'s to the diagonal subgroup.

Similar remarks apply to $H'$ spinor indices, except the range of indices decreases by factors of 2, and the ovals are opposite pairings, so as to produce a total of 32 real $Î$'s.  Also, the groups correspond to $L$ (except for range), while the index structure of $Y$ matches that of $§$ because of how $\di$ appears in the algebra.  Thus $Y$ is a bispinor of $H'$, while $Î$ is a spinor of both $H$ and $H'$, and $\widetilde Y$ is a bispinor of both.

So, for example, for $\mathrm D=6$ (where the Lorentz group is $SU*(4)$) we have that $H=SU*(8)$ and $L=SU*(8)^2$, while $H'=USp(2)^2$ ({\it i.e.\ }$SO(4)$ for $\mathrm D'=10- \mathrm D=4$), with $(X^{[Œº]},X_{[Ќк]})$, $§^Œ{}_{к}$, $Y^{Œ'}{}_{к'}$, and $(Î^{ŒŒ'},Î_{ЌЌ'})$.  (Unlike $\mathrm D<6$, for this case there is an additional {\bf 70} of $§$ not apparent from $ÓD,DÕ$.  For $\mathrm D=7$ there is furthermore a {\bf 128} of $X$ not apparent from $ÓD,DÕ$.)

%%%%%%%%%%%%%%%%%%%%%%%%%%%%%%%%%%%%%%%%%%%%%%%%%%%%%%%%%%%%%%%%
\section{\texorpdfstring{$ú$}{\texteta}'s}
\label{S:eta}
The fundamental tensor from which all others are constructed is the $E_n$ Clebsch-Gordan-Wigner coefficients $\eta$ mapping the symmetric product of the representation $R_n$ with itself to $R_1$. (The numbering of the fundamental representations is as indicated in the top row of fig. \ref{F:Fdiamond}.) When $n\leq 5$ the exceptional isomorphisms give these tensors in terms of classical Lie algebra invariants. 
We collect the representations and $\eta$ tensors in table \ref{T:EReps}.

%\begin{table}[ht]
%\begin{align*}
%{\renewcommand{\arraystretch}{1.3} %adds some padding
%\begin{array}{c|ccccccc}
%\mathrm D &1 & 2 & 3& 4& 5& 6 \\
%\hline
%E_{n(n)} & SL(2)   & SL(3)\times SL(2)  & SL(5) & Spin(5,5) &E_{6(6)}&E_{7(7)}\\
%\sigma^a  & \bm2 &(\bm3, \bm1) &\bm 5&\bm{10}& \bm{27}& \bm{133} \\
%P_A & \bm1\oplus\bm2 & (\bm3, \bm2) & \bm{10} 	& \bm{16} & \bm{27}& \bm{56} \\
%%\sigma^a  & \bm2 &(\bm3, \bm1) &\bm 5&\bm{10}& \bm{27}& (AB)\in \bm{133} \\
%%P_A & (1,a) \in \bm1\oplus\bm2 & ai\in (\bm3, \bm2) & [ab]\in \bm{10} 
%%	& A\in \bm{16} & a\in \bm{27}& \bm{56} \\
%a \on?\sim A & A=1,a & A=ai & A=[ab] & & A=a & (AB) \ni a\\
%\eta_{ABc}  & \epsilon_{ab}&\epsilon_{abc} C_{ij} & \epsilon_{[ab][cd]e} 
%	& (\gamma_c)_{AB} & d_{abc} & d_{(AB)CD}
%\end{array}
%}
%\end{align*}
%\caption{\small The $E_{\mathrm D+1}$ groups corresponding to the D-dimensional strings, the representations of the worldvolume (without $\tau$) and target coordinates, and the Clebsch-Gordan-Wigner tensors relating them.
%{\color{red} The $\eta$ for $E_7$ seems wrong. How can there not be a CGW for $\bm{56} \otimes \bm{56} \to \bm{133}$ ??}
%}
%\label{T:EReps}
%\end{table}
\begin{table}[ht]
\begin{align*}
{\renewcommand{\arraystretch}{1.3} %adds some padding
\begin{array}{c|ccccccc}
\mathrm D &1 & 2 & 3& 4& 5\\
\hline
E_{n(n)} & SL(2)   & SL(3)\times SL(2)  & SL(5) & Spin(5,5) &E_{6(6)}\\
\sigma^a  & \bm2 &(\bm3, \bm1) &\bm 5&\bm{10}& \bm{27}\\
P_A & \bm1\oplus\bm2 & (\bm3, \bm2) & \bm{10} 	& \bm{16} & \bm{27}\\
a \on?\sim A & A=1,a & A=ai & A=[ab] & & A=a \\
\eta_{ABc}  & \epsilon_{ab}&\epsilon_{abc} C_{ij} & \epsilon_{[ab][cd]e} 
	& (\gamma_c)_{AB} & d_{abc} 
\end{array}
}
\end{align*}
\caption{\small The $E_{\mathrm D+1}$ groups corresponding to the D-dimensional strings, the representations of the worldvolume (without $\tau$) and target coordinates, and the Clebsch-Gordan-Wigner tensors relating them.
}
\label{T:EReps}
\end{table}

Note that the case $\mathrm D=1$ corresponds to a scalar $X$ and its dual $\widetilde X_a$. 
Thus, the ``internal'' scalars $Y^{a'}$ and their duals $\widetilde Y_{aa'}$ are a special case of the general construction in section \ref{S:CurrentAlgebra}: We replace the selfdual indices $A\to A' = \un a a'$ (so $Y^{A'} = Y^{\un a a'} = (Y^{a'},\widetilde Y^{aa'})$) and take 
\begin{align}
\eta_{A'B'c} %= (\eta_{\un aa' \, \un bb'})_c 
= \left(
	\begin{array}{cc}
	0& \delta_{a'b'} \eta_{bc}\\
	\delta_{a'b'} \eta_{ac}& 0\\
	\end{array}
\right) 
\end{align}
The $O({\mathrm D}',{\mathrm D}')$-like invariant can, itself, be extended by the $\tau$-component 
\begin{align}
\eta_{A'B' 0} = \delta_{a'b'} \eta_{\un a\un b} .
\end{align}
With this, the construction of the contributions of the scalars to the Virasoro, Gau\ss{} law, Laplace, {\it et cetera} constraints is the same as that given for the $X$'s in section \ref{S:CurrentAlgebra}.

%%%%%%%%%%%%%%%%%%%%%%%%%%%%%%%%%%%%%%%%%%%%%%%%%%%%%%%%%%%%%%%%
%%%%%%%%%%%%%%%%%%%%%%%%%%%%%%%%%%%%%%%%%%%%%%%%%%%%%%%%%%%%%%%%
{
%\small
\footnotesize
\linespread{1.1}\selectfont
\raggedright
\bibliography{/Users/wdlinch3/Dropbox/Rashoumon/LaTeX/BibTex/BibTex}
\bibliographystyle{unsrt}
} % end resize

\end{document}